\documentclass[prl,english,twocolumn,amsmath,amssymb,superscriptaddress]{revtex4-1}
\usepackage{amsmath}
\usepackage{graphicx}
\usepackage{amssymb}
\usepackage{dsfont}
\usepackage{color}
\usepackage[bookmarks=true,colorlinks,linkcolor=blue,urlcolor=blue,citecolor=blue]{hyperref}
\usepackage{hyperref}
\usepackage{tikz}
\usepackage{soul}
\usetikzlibrary{calc}

\newcommand{\be}{\begin{equation}}
\newcommand{\ee}{\end{equation}}
\newcommand{\bea}{\begin{eqnarray}}
\newcommand{\eea}{\end{eqnarray}}

\newcommand{\mc}{\mathcal}

\begin{document}
	
\title{Nonlinear effects on   charge fractionalization in critical chains}

	\author{Fl\'avia B. Ramos}
	\affiliation{Physics Department and Research Center OPTIMAS, University of Kaiserslautern-Landau, 67663 Kaiserslautern, Germany}
	\author{Rodrigo G. Pereira}
\affiliation{International Institute of Physics and Departamento de F\'isica Te\'orica e Experimental, Universidade Federal do Rio Grande do Norte, Natal, RN, 59078-970, Brazil}
		\author{Sebastian Eggert}
	\affiliation{Physics Department and Research Center OPTIMAS, University of Kaiserslautern-Landau, 67663 Kaiserslautern, Germany}
	\author{Imke Schneider}
	\affiliation{Physics Department and Research Center OPTIMAS, University of Kaiserslautern-Landau, 67663 Kaiserslautern, Germany}
	
	\begin{abstract}

We investigate the generic transport in a one-dimensional 
strongly correlated fermionic chain beyond linear response.    
Starting from a Gaussian wave packet  with  positive momentum on top of the ground state, we find that the  numerical time evolution splits the signal
into at least three distinct fractional charges moving with different velocities.  
A fractional left-moving charge is expected from conventional Luttinger liquid theory, but 
for the prediction of the two separate right-moving packets the nonlinearity of the dispersion
must also be taken into account. This out-of-equilibrium protocol therefore 
allows a  direct measurement of nonlinear interaction parameters, which also
 govern threshold singularities of dynamic response functions. The 
nonlinear Luttinger Liquid theory also predicts the correct dynamics at low energies, where it agrees
with the conventional Luttinger liquid.
Moreover, at high energies, the wave packet dynamics reveals signatures of composite excitations containing two-particle bound states.  
Our results uncover a  simple strategy  to probe the nonlinear regime  
in time-resolved  experiments in quantum wires and ultracold-atom platforms.
\end{abstract}

\date\today

\maketitle

\emph{Introduction.---} The fractionalization of a particle into a composite of emergent excitations  is  one of the most striking phenomena in quantum many-body systems. The effect is  prevalent in critical one-dimensional (1D)  systems, in which interactions inevitably lead to a  departure from Fermi liquid behavior \cite{Giamarchi,Haldane1981,Deshpande2010}. Within the paradigm of  Luttinger liquid (LL) theory \cite{Tomonaga1950,Luttinger1963}, the low-energy spectrum of      1D quantum fluids  is  described by   bosonic collective modes with a linear  dispersion relation. This theory  predicts that electrons  fractionalize into right- and left-moving  excitations that carry interaction-dependent charges \cite{Pahm2000,Leinaas2009}, as indeed  observed   in transport experiments in quantum wires \cite{Steinberg2007,LEHUR20083037,Kamata2014,Freulon2015}. In addition, signatures of LL behavior have been identified via spectroscopic techniques   \cite{Kim1996,Segovia1999,Auslaender2002,Auslaender2005,kim2006,Mourigal2013} and quantum simulations in ultracold-atom platforms \cite{Kinoshita2004,Paredes2004,Pagano2014,Hilker2017}.

Despite its impressive success, LL theory breaks down whenever finite-energy excitations and band curvature  have to be  taken into account \cite{ImambekovRMP2012}. To treat the effects of a nonlinear dispersion,  more general techniques have been developed into what became known as the nonlinear Luttinger liquid (nLL) theory  \cite{Rozhkov2005,Pustilnik2006,Khodas2007,Imambekov2009,Pereira2008}. In particular, dynamic response functions exhibit characteristic threshold singularities that can be described  by treating the    modes with finite energy and momentum  as    mobile impurities coupled to the gapless modes \cite{Pustilnik2006}. The exponents associated with these singularities can be expressed in terms of scattering phase shifts and calculated exactly for integrable models \cite{Pereira2008,Cheianov2008,Imambekov2008,Essler2010}. In the time domain, the contributions from  high-energy modes give rise to power-law-decaying temporal oscillations  that   dominate the long-time behavior \cite{Pereirareview}.  Moreover,   nonlinearities are predicted to lead  to shock waves in the evolution of density and magnetization pulses \cite{Bettelhein2006,Bettelheim2012,Protopopov2013,Protopopov2014}.

While the nonlinear regime is accessible in experiments \cite{Barak2010,Jin2019,Wang2020,Senaratne2022},  direct  tests of the  threshold singularities predicted by nLL theory    are hindered by the limited energy resolution or disorder-induced broadening of spectroscopic probes. In this work, we show that the effects of the high-energy excitation in   nLL theory can be directly observed in an out-of-equilibrium protocol. We create a Gaussian wave packet   with preselected  momentum in a critical fermionic chain and   simulate its evolution using the adaptive time-dependent density matrix renormalization group (tDMRG) \cite{white2004}.  Similar protocols have been used to demonstrate   fractionalization and spin-charge separation in the low-energy regime  \cite{Hallberg1993,PhysRevLett.92.226405,Kollath2005,Ulbricht_2009,Hassanieh2013,Acciai2017,Scopa2021}. Beyond the LL paradigm, the nonlinear dispersion   leads to a splitting of  the initial wave packet   into three   density humps that propagate with different velocities \cite{Moreno2013,Dontsov2021}. Here, we show that the time-evolved signal can be predicted by
nLL theory, which in turn provides a quantitative measurement of the 
interaction between the high-energy particle and the low-energy modes. At lower fillings and higher energies, we discover fingerprints of  two-particle bound states in the wave packet dynamics.

\emph{Model and protocol.---} We consider a spinless fermion model described by the   Hamiltonian
\begin{equation}
    H=\sum_{j=1}^L\left[-\frac{1}{2}\left(c^\dagger_jc^{\phantom\dagger}_{j+1}+\text{h.c.}\right)+V n_j n_{j+1} \right],\label{Ham}
\end{equation}
where $c_j$ annihilates a fermion at site $j$ of a chain with  $L$ sites, $V$ is the strength of the nearest-neighbor   interaction, and $n_j=c^\dagger_jc^{\phantom\dagger}_j$ is the local density operator. We work at fixed   number  of fermions $N$, with average density $n=N/L$. For $V=0$,  the   Hamiltonian can be diagonalized as $H_0=\sum_{k}\varepsilon_0(k)c_k^\dagger c^{\phantom\dagger}_k$, where $\varepsilon_0(k)=-\cos (k)$ is the dispersion relation of free fermions with momentum $k\in [-\pi,\pi]$, with lattice spacing set to unity. The ground state in this case is constructed by occupying single-particle states up to the Fermi momentum $k_F=\pi n$. More generally, the model in Eq. (\ref{Ham}) is equivalent to the spin-1/2 XXZ  chain and is exactly solvable via Bethe ansatz  (BA) \cite{Korepin1993}.  We focus on the parameter regime $0\leq V\leq1$, where the ground state $\left |\Psi_0\right\rangle$ of Eq. (\ref{Ham}) is in a gapless phase described at low energies by   LL theory \cite{Giamarchi}.   

\begin{figure}[t]
    \includegraphics[width=0.95\columnwidth]{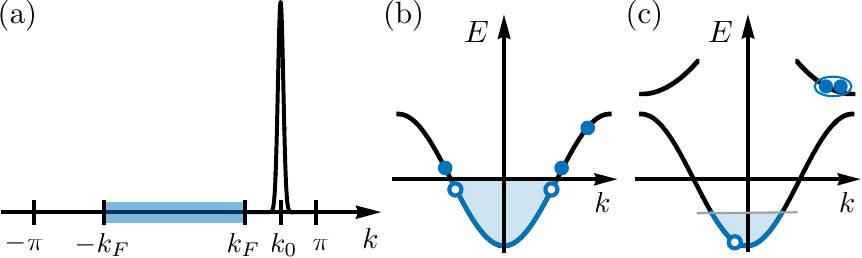}
    \caption{(a)  Gaussian wave packet in momentum space. (b) At half filling, the evolution is controlled by a   particle above the Fermi level and low-energy particle-hole excitations. (c) At lower fillings, there are contributions from   excitations with a hole and two particles in a  high-energy bound-state band. \label{fig1}}
\end{figure}

We prepare an  initial state  given by \be
\left|\Psi(t=0)\right\rangle=  A\sum_{j=1}^{L}e^{-\frac{(j-j_0)^2}{2\sigma_0^2}}e^{ik_0 j}c^\dagger_j\left|\Psi_0\right\rangle,\label{exclat}
\ee
corresponding to a Gaussian wave packet centered at $j_0$, with variance $\sigma_0^2/2$ in real space and mean momentum $k_0$; here, $A$ is a normalization constant. To add a particle   with well-defined momentum, we choose $k_0\in[k_F,\pi]$ with momentum uncertainty $\Delta k=\frac1{\sqrt2\sigma_0}$; see Fig. \ref{fig1}(a). After the unitary evolution $\left| \Psi(t)\right\rangle =e^{-iHt}\left| \Psi(0)\right\rangle$, we   measure the time-dependent local charge excess defined as \be
 \rho_j(t)=\langle \Psi(t)|n^{\phantom\dagger}_j|\Psi(t)\rangle-\langle \Psi_0|n^{\phantom\dagger}_j|\Psi_0\rangle.\label{tdens}
\ee

\emph{Free fermions.---} The local charge excess can be calculated exactly in the noninteracting case. For a finite chain with open boundary conditions, we obtain \be
\rho_j(t)=     \left|\sum_{l=N+1}^Lq(k_l)\sin(k_l j)  e^{i \cos(k_l) t}\right|^2 ,
\label{freefermion}
\ee
where  $q(k)=\frac{2A}{L+1}\sum_{j=1}^L  e^{-\frac{(j-j_0)^2}{2\sigma_0^2}}e^{ik_0 j}\sin(kj)$ and $k_l=\frac{l \pi}{L+1}$ with $l=1,\cdots,L$.  The result is analogous to the evolution of a Gaussian wave packet in the single-particle problem \cite{Shankar1994}. For small $\Delta k$, we can expand the dispersion    as $\varepsilon_0(k)\approx \varepsilon_0(k_0)+u_0(k-k_0)+\frac1{2m_0}(k-k_0)^2$, where $u_0=\sin( k_0)$ is the group velocity and $m_0=[\cos (k_0)]^{-1}$ is the effective mass for momentum $k_0$. As a consequence, we observe a single packet that moves to the right with velocity $u_0$ and width growing as 

\be
\sigma(t)^2\approx {\sigma_0^2} \left(1+\frac{t^2}{m_0^2\sigma_0^4}\right).\label{width}
\ee

We have used Eq. (\ref{freefermion}) to benchmark our tDMRG results, obtaining excellent  agreement; see the Supplemental Material (SM) \cite{SM}.  In all the numerics henceforth, we set $L=300$, $j_0=150$, and $\sigma_0=14.5$. We keep up to 400 states per DMRG block and use the Trotter step $\delta t=0.1$. The largest truncation error   is of order $10^{-10}$. The maximum time is set by stopping the simulations before the wave packets reach the chain boundaries. 

\emph{nLL theory.---}We now turn to  the interacting case. We seek to  describe the dynamics  using  the framework of    nLL theory \cite{ImambekovRMP2012}. In addition to the   momentum $k_0$, we  consider a mode expansion that includes the low-energy modes in the vicinity of the Fermi points [see Fig.~\ref{fig1}(b)]:
\begin{equation}
   c^\dagger_j \sim e^{-ik_Fx}\psi^\dagger_R(x)+e^{ik_Fx}\psi^\dagger_L(x)+e^{-ik_0x}d^\dagger(x).\label{modexp}
\end{equation}
In the absence of the high-energy particle created by $d^\dagger$, the low-energy modes are described by the LL Hamiltonian  \cite{Giamarchi}\be
H_{\rm LL}=\frac{v}{2}\int dx\,\left[\left(\partial_x\varphi_R\right)^2+\left(\partial_x\varphi_L\right)^2\right],
\ee
where $v$ is the velocity of the low-energy modes and $\varphi_{R,L}(x)$ are the right- and  left-moving components of the bosonic field    obeying 
\be[\partial_x\varphi_{R,L}(x),\varphi_{R,L}(x')]=\pm i{\delta(x-x')}.\label{comm}\ee   The low-energy   fermion fields can be bosonized in the form   \be
\psi^\dagger_{R,L}(x)\sim e^{i\sqrt{\frac{\pi}{2K}}\left[(1\pm K)\varphi_R(x)+(1\mp K)\varphi_L(x)\right]},\label{psiR}
\ee
where $K$ is the Luttinger parameter. Taking   the continuum limit of Eq. (\ref{Ham}) including the high-energy mode, we obtain the  effective Hamiltonian
\begin{align}
H_{\rm nLL}&=H_{\rm LL}+\int dx\, d^\dagger (\varepsilon_{\rm p}-i u\partial_x )d
\nonumber \\
&    +\frac{1}{\sqrt{2\pi K}}\int dx\, \left(\kappa_R\partial_x\varphi_R+\kappa_L\partial_x\varphi_L\right)d^\dagger d ,
\label{himp} 
\end{align}
where $\varepsilon_{\rm p}$ and $u$ are the renormalized energy and velocity, respectively, of the high-energy particle in the interacting model. Note that the high-energy particle behaves as a mobile impurity \cite{Tsukamoto1998,Balents2000} that interacts with the   bosonic  modes     via the coupling constants $\kappa_{R,L}$.

 The impurity mode can be decoupled from the bosonic fields by the unitary transformation $
	U=e^{-{i}\int\frac{dx}{\sqrt{2\pi K}}\left(\gamma_R \varphi_R+\gamma_L\varphi_L\right)d^\dagger d}$,
	where $\gamma_{R,L}=\kappa_{R,L}/(v\mp u)$  are the  right- and left-mover phase shifts. In the nLL, $\gamma_{R,L}$ govern all the space-time correlation functions of the system as well as the threshold singularities of dynamic response functions \cite{Pustilnik2006,Pereira2008}. 
As we will see below, these non-trivial interaction parameters can be measured as fractional charges in our proposed out-of-equilibrium  protocol.

We now proceed to calculate the fractional transport properties using the nLL theory.
Up to irrelevant terms, the transformed Hamiltonian $\tilde{H}_{\rm nLL}=U  {H}_{\rm nLL}U^\dagger$ becomes noninteracting \cite{Pereira2009}
\begin{equation}
\tilde{ {H}}_{\rm nLL}= \frac{v}{2} \left[\left(\partial_x \tilde\varphi_{R}\right)^2+\left(\partial_x \tilde\varphi_{L}\right)^2\right]+\tilde d^\dagger (\varepsilon_{\rm p}-iu\partial_x)\tilde d,\label{freeham}
\end{equation}
with 
\bea
\partial_x \varphi_{R,L} &  = &  U^\dagger\partial_x \tilde\varphi_{R,L}U = \partial_x \tilde\varphi_{R,L}\pm \frac{\gamma_{R,L}}{\sqrt{2 \pi K}} \tilde d^\dagger \tilde d \label{phi},\\
d^\dagger & = &  U^\dagger \tilde d^\dagger U= \tilde d^\dagger\,  e^{\frac{i}{\sqrt{2\pi K}}\gamma_R\tilde\varphi_R}e^{
\frac{i}{\sqrt{2\pi K}}\gamma_L \tilde\varphi_{L}}.\label{hepart}
\eea

Note that an initial excitation $d^\dagger$ now consists of 
three parts: 
 two vertex operators, $\mc V_{L,R}(x)=e^{\frac{i}{\sqrt{2\pi K}}\gamma_{L,R}\tilde\varphi_{L,R}(x)}$, that excite low-energy modes $\tilde\varphi_{L,R}(x,t)=\tilde\varphi_{L,R}(x\pm vt)$ 
propagating with velocity $\pm v$ and a free particle $\tilde d^\dagger$ with velocity $u$, as schematically depicted in Fig.~\ref{fig1}(b).
The fractional charges of the three propagating parts can be measured with 
the charge density operator, which after rotation using Eq.~(\ref{phi}) is given by
\bea
\hspace{-.4cm}
Q= \sqrt{\frac{K}{2 \pi}}\partial_x(\tilde \varphi_L-\tilde\varphi_R) +\left(1-\frac{\gamma_R+\gamma_L}{2\pi}\right)\tilde d^\dagger \tilde d.
\eea
By calculating the commutator  using Eq.~(\ref{comm}), we have
\be [Q(y),\mc V_{R,L}(x)]=\frac{\gamma_{R,L}}{2 \pi}\mc V_{R,L}(x)\delta(x-y).
\ee
Thus, the right- and left-moving vertex operators carry fractional charges $n_{R,L}=\frac{\gamma_{R,L}}{2 \pi}$ propagating with velocity $\pm v$. The  remaining charge corresponding to the particle $\tilde d^\dagger$ is  $n_I=1-n_R-n_L$ with velocity $u$.

\begin{figure}
    \centering
\includegraphics[width=7cm]{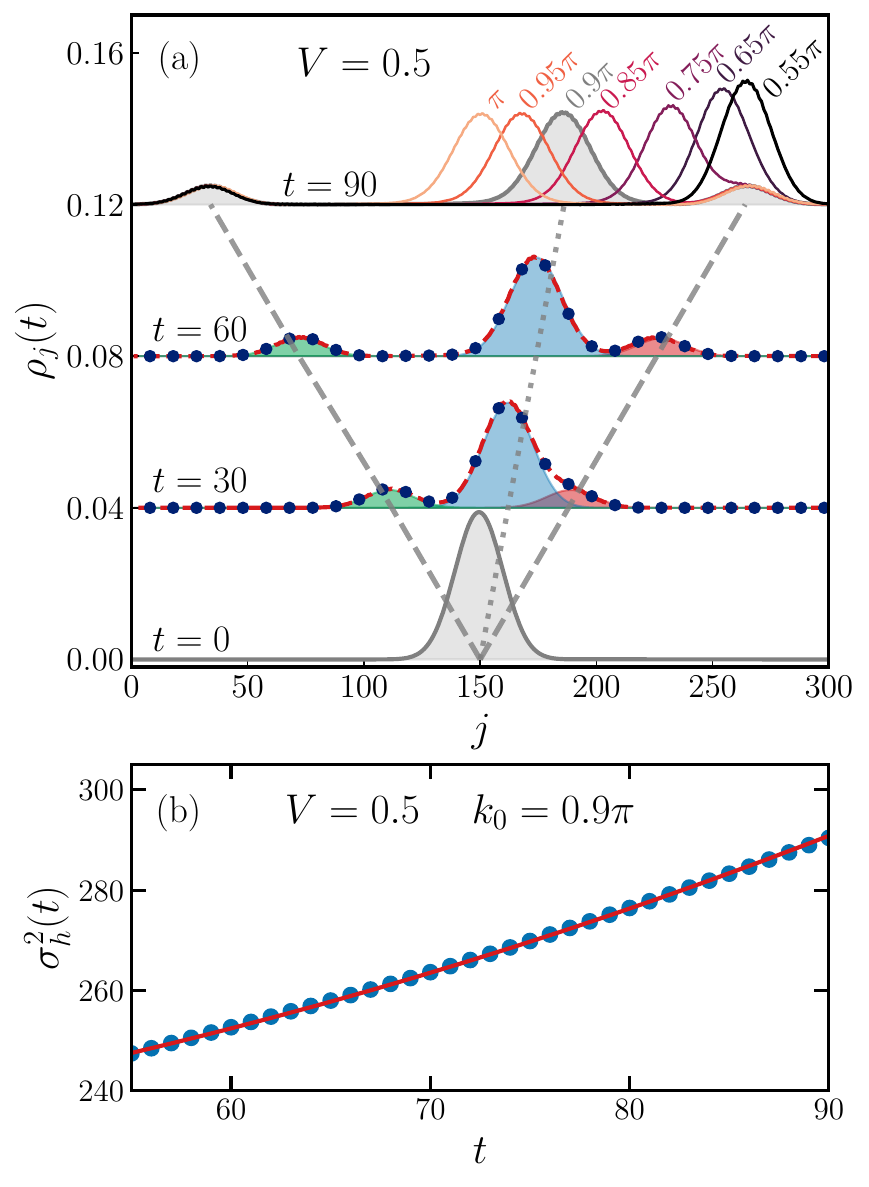}
    \caption{Wave packet   at half filling. (a) Snapshots of the averaged density profile for different times and $V=0.5$. The time evolution   is shown for $k_0=0.9\pi$. The gray dashed lines denote the light cone set by the velocity of the low-energy modes. 
    The dotted line indicates the   velocity of the high-energy particle. The shaded regions for the data $t=30$ and $60$ show the time-dependent  Gaussian functions which reproduce the tDMRG results represented by the symbols. The red dashed lines are the  sums of the Gaussian functions.  For $t=90$, we show density profiles for the values of $k_0$ indicated in the plot. (b) Variance of the high-energy hump vs. $t$. The red solid line is a fit to our estimates using a quadratic function \cite{SM}.}
    \label{fig2}
\end{figure}

In the tDMRG simulation, we use the Gaussian wave packet defined in Eq.~(\ref{exclat}); see Fig.~\ref{fig1}(a).
The propagation at half-filling, $k_F=\pi/2$, is shown in Fig.~\ref{fig2}(a). Note that all parameters are known in this case: $K=\frac{\pi}{2[\pi - \arccos(V)]}$, $v=\frac{\pi\sqrt{1-V^2}}{2\arccos(V)}=u/\sin(k_0)$, and \emph{momentum-independent} phase shifts $\gamma_{R,L}=\pi(1-K)$ \cite{Pereira2008}. In all cases, we observe three fractional humps that move with the exact velocities  and magnitudes predicted by the nLL theory. Remarkably, the shape of the low-energy wave packets  is stable over the whole energy regime, showing coherence over long times as it can be seen in Fig. \ref{fig2}(a). In addition, the variance of the high-energy hump, $\sigma_h^2$, grows in time according to Eq. (\ref{width}) with an interaction-dependent effective mass; see Fig. \ref{fig2}(b). Therefore, the theoretical ad-hoc prediction of exactly three parts in the mode expansion (\ref{modexp}) surprisingly provides a robust prediction of a free stable fractional particle. See the SM \cite{SM} for more details on the fitting procedure.

\begin{figure*}[t]
	\centering
	\includegraphics[width=0.81\paperwidth]{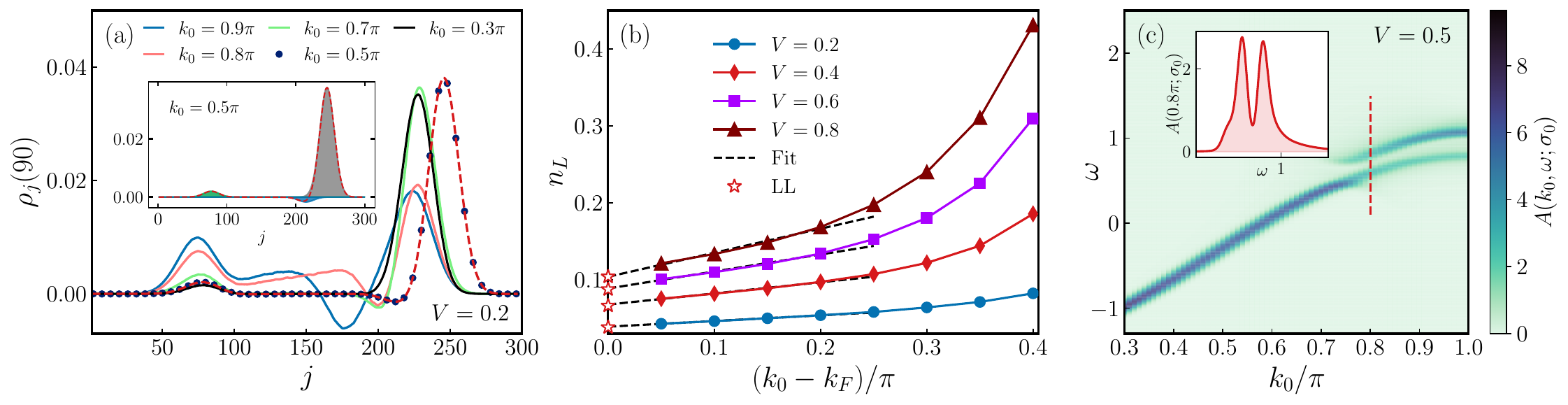}
	\caption{Wave packet  at quarter filling.   (a) Averaged density profile for at fixed time $t=90$, $V=0.2$, and distinct values of $k_0$. For $k_0=0.5\pi$, the symbols are the tDMRG results and the dashed line is the fit to our data considering the sum of three Gaussian functions, which are shown in the inset.
		(b) Charge of the left-moving hump as a function of momentum for different values of $V$.  The dashed lines represent a  linear fit of the data for $(k_0-k_F)/\pi\leq 0.2$. The stars on the vertical axis mark  the LL prediction $(1-K)/2$, valid for $k_0\to k_F$. (c) Spectral function  $A(k_0,\omega;\sigma_0)$, see Eq. (\ref{respf}), calculated for $V=0.5$. The inset shows a cut for momentum $k_0=0.8\pi$. The   peak at higher frequencies is a signature of bound states.  
	}
	\label{fig3}
\end{figure*}

Here, spatial oscillations in the density with wave number  $2k_F$ have been averaged out by  taking the average $\rho_j(t)$ between two nearest-neighbor sites.
They can  be attributed to  the staggered part of the density operator as discussed in the SM \cite{SM}.

In our analysis, we have used a large set of different values of $t$, $V$, and  $k_0$, which all show excellent agreement with the nLL theory.  
This implies that the nLL theory remarkably describes the \emph{whole} energy regime covered by the Gaussian excitation with momentum $k_0>k_F$. As predicted, we have not seen any momentum dependence in $n_{R,L}$ \cite{SM}. Moreover, we observe a left-right symmetry in the low-energy humps, which is surprising since the initial wave packet at finite $k_0$ clearly does not have this symmetry.  Last but not least, we can apply the theory also to excitations very close to the Fermi energy, i.e. $k_0\to \pi/2$.  Once $u\to v$, longer and longer times
	$t\agt \tau_{\rm sp}\equiv  \sigma_0/|v-u|$ are required to distinguish the two right-moving humps. In fact, this is in perfect agreement 
	with conventional LL theory \cite{Pahm2000} of only one left- and one right-moving 
	fractional charge with
	$\frac{1-K}2$  and $\frac{1+K}2$, respectively \cite{Pahm2000}.
 Hence, the continuous crossover from nLL to LL behavior becomes very intuitive in the dynamics of  fractional 
	charges $\frac{\gamma_{L,R}}{2 \pi}$. In contrast, this crossover is much more involved in the frequency domain since the threshold exponents are quadratic functions of the phase shifts \cite{Imambekov2009}.

\emph{Quarter filling.---}To illustrate a more generic situation, we now consider the model in Eq. (\ref{Ham}) at quarter filling, $k_F=\pi/4$. In this case, by solving Bethe ansatz equations, we can numerically determine the  exact renormalized dispersion, Luttinger parameter, and velocity of the low-energy modes \cite{Giamarchi,Korepin1993}.  

Figure  \ref{fig3}(a) shows  tDMRG  results for $V=0.2$ and different values of $k_0$ at $t=90$ after averaging  
the density over four neighboring sites to smooth the $2k_F$ oscillations out.  
In the low-energy regime, e.g.~for $k_0=0.3\pi$,
we observe only one left and one right hump moving with velocities  $\pm v$, similar to the half-filled case. 
However, now the charge $n_L$ carried by the left-moving hump varies with $k_0$ as shown in  Fig. \ref{fig3}(b), which in turn directly provides the momentum-dependent phase shifts $\gamma_L=2\pi n_L$. In the extrapolation $k_0\to k_F$ we recover the LL prediction $n_L\to \frac{1-K}{2}$. 

For larger values of $k_0$, see for example  $k_0=0.5\pi$ and $0.7\pi$ in Fig. \ref{fig3}(a), the three signals predicted by the nLL are observed again: two counter-propagating charges $n_{R,L}=\frac{\kappa_{R,L}}{2\pi(v\mp u)}$ moving with velocities $\pm v$, and a large right-moving hump with velocity $u$. Note that the right low-energy mover carries \emph{negative} charge since  $u>v$. We again find that the time evolution of the excitation can be predicted by three propagating Gaussian wave packets \cite{SM}.
 
As we increase the momentum, the density profile develops a complex pattern. 
For  $k_0=0.8\pi$ in Fig. \ref{fig3}(a),  the broad feature around the middle of the   chain  suggests the presence of additional excitations. In fact,   the   operator in Eq.~(\ref{exclat}) can create not only individual particles but also  composite excitations with net charge +1.   To analyze the excitations, we consider the Fourier transform  of the overlap between the  initial state in Eq.~(\ref{exclat}) and the time-evolved one 
\begin{align}
     A(k_0,\omega;\sigma_0)=& \int_{-\infty}^\infty dt\,  e^{i(\omega-E_0) t} \left\langle \Psi(0)|\Psi(t)\right\rangle \nonumber, \\
       = &  2\pi\sum_\alpha |\langle\alpha|\Psi(0)\rangle|^2\delta (\omega - E_\alpha + E_0),\label{respf}
\end{align}
where $|\alpha\rangle$ are eigenstates of $H$ with energy $E_\alpha$ above the ground-state energy $E_0$.
In the limit $\sigma_0\rightarrow \infty$,   it  reduces to the   standard single-particle spectral function  \cite{Pereira2009}. 
We compute $ A(k_0,\omega;\sigma_0)$ by performing a fast Fourier transform with a cosine window in the interval $t\in[-t_\text{max},t_\text{max}]$, with $t_\text{max}=90$. 

As discussed in Ref.~[\onlinecite{Pereira2009}],
the nLL theory for $ A(k_0,\omega,\sigma_0\to \infty)$ predicts a diverging peak  at 
the dispersion $\omega=\varepsilon_\text{p}$,  which decays 
as power laws on both sides with exponents dependent on the phase shifts.  In our results  of  $A(k_0,\omega;\sigma_0)$, this is reflected as a broad single-particle peak for $k_0 \alt 0.6 \pi$; see Fig. \ref{fig3}(c).  
At larger momenta, $A(k_0,\omega;\sigma_0)$
displays a double-peak structure for $k_0$ near $\pi$, which is a clear indication of
a yet unaccounted excitation in the 
wave packet $|\Psi(0)\rangle$ in Eq.~(\ref{specd}).
This feature is absent for half-filling \cite{SM}.

We identify this excitation as a composite of a
two-particle bound state with a free hole as depicted in Fig.~\ref{fig1}(c), 
which has been predicted in Ref.~\cite{Pereira2009}. Such a composite gives a high-energy continuum in the spectral function for momenta greater than $\pi + k_F - Q_{\text{bs}}$, where  $Q_{\text{bs}}=\left[\pi- 2\arccos(V)\right]\left(1-\frac{2k_F}{\pi}\right)$ restricts the momentum of bound state $|\pi-k_{\text{bs}}|<Q_{\text{bs}}$.
Note that $Q_{\rm bs}$ vanishes for $V\to 0$ or $k_F\to \frac\pi2$.  
In nLL theory, this type of excitation can be described by   $c_j^\dagger\sim e^{-ik_0x}B^\dagger(x)h^\dagger(x)$ \cite{Pereira2009}.  For  $k_0=0.8\pi$ in Fig.~\ref{fig3}(a), both the bound state ($B^\dagger$) and the hole ($h^\dagger$) are  created near the bottom of their respective  bands.  This implies a low velocity,  consistent with a  slow-moving  hump. In general, the evolution  of the wave packet may involve  contributions from low- or high-energy particles, holes, and bound states that share the total momentum covered by the Gaussian distribution; see the SM \cite{SM}. 

\emph{Conclusions.---} We proposed an out-of-equilibrium protocol to investigate the fractionalization of   high-energy excitations    in critical chains. We clarified the crossover between LL and nLL regimes  and   identified  contributions from  elementary  particles 
moving with different velocities and carrying fractional charges, which can be negative  
for $u>v$.  The transport simulations reveal also more involved excitations, such  composite exctiations formed by holes and bound states.
This analysis also applies to quantum spin chains and spin-charge separated quantum fluids \cite{SchmidtPRL2010,ImkeEsslerRodrigo}. 
Our work paves the way for precision tests of nLL effects  through non-equilibrium dynamics in  ultracold-atom platforms  \cite{Pedersen2013,Vijayan2020} and  time-resolved measurements of hot electrons   in quantum wires  and quantum Hall edge states \cite{ Kataoka2016,Hashisaka2017,Bauerle2018}. 
In particular, we  find a fractionally charged particle with free-particle dynamics, a right-moving low-energy excitation which can be negatively charged, and a left-moving signal that gives an exact measurement of the interaction couplings in a large  parameter regime.
Hence, experimental measurements of counter-propagating fractional charges 
directly provide quantitative values of the momentum-dependent
interactions in the linear and the nonlinear regimes.

\begin{acknowledgments}
This work was supported by the 
Deutsche Forschungsgemeinschaft (DFG, German Research Foundation) - Project 
No. 277625399-TRR 185 OSCAR (A4,A5), 
by the Conselho Nacional de Desenvolvimento Cient\'{i}fico e Tecnol\'ogico (R.G.P.), and  by a grant from the Simons
Foundation (Grant No. 1023171, R.G.P.). The authors thank the high-performance cluster Elwetritsch for providing computational resources.
\end{acknowledgments}

\bibliographystyle{apsrev4-1_control}

\pagebreak
\section*{Supplemental material}

\subsection*{Free-fermion case}

\begin{figure}[b]
	\begin{centering}
		\includegraphics[width=7cm]{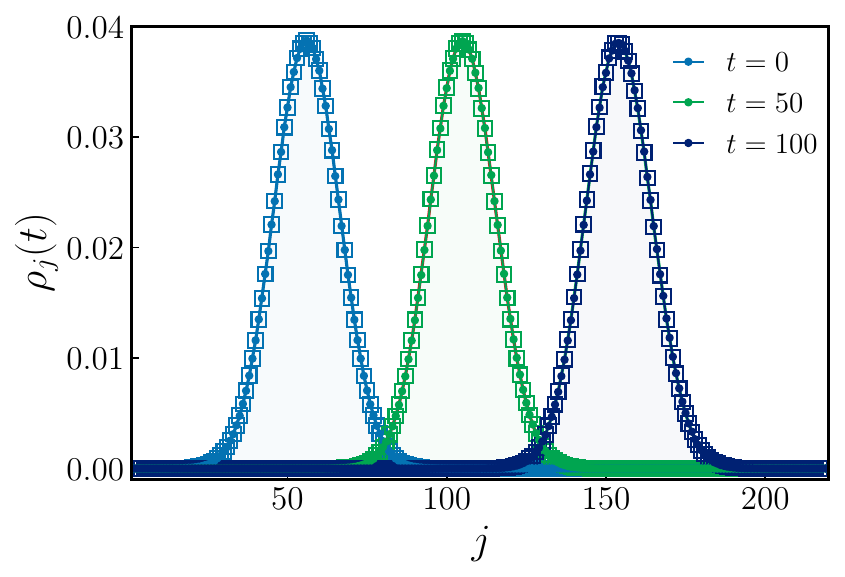}
		\par\end{centering}
	\caption{Snapshots of the density profile for $k_{0}=0.56\pi$ and half
		filling in the noninteracting model, $V=0$.
		The filled circles are the exact results obtained via Eq. (\ref{freefermion}) of the main text and the
		squares are tDMRG data. The results were obtained setting $L=220$,
		$\sigma_0=14.5$, and $j_{0}=55$.\label{fig:ffdmrg}}
\end{figure}

We use the exact solution of the free-fermion case to benchmark our
tDMRG results. In Fig. \ref{fig:ffdmrg}, we show snapshots of typical
density profiles for half filling. We find an excellent
agreement between the analytical formula 
and the numerical simulation. Due to the relatively small variance
in the Fourier space, all the one-particle states with significant
probability amplitude have roughly the same velocity. As a consequence,
we observe the coherent motion of the wave packet.

\subsection*{Oscillations in the density profile}

\begin{figure}
	\begin{centering}
		\includegraphics[width=7cm]{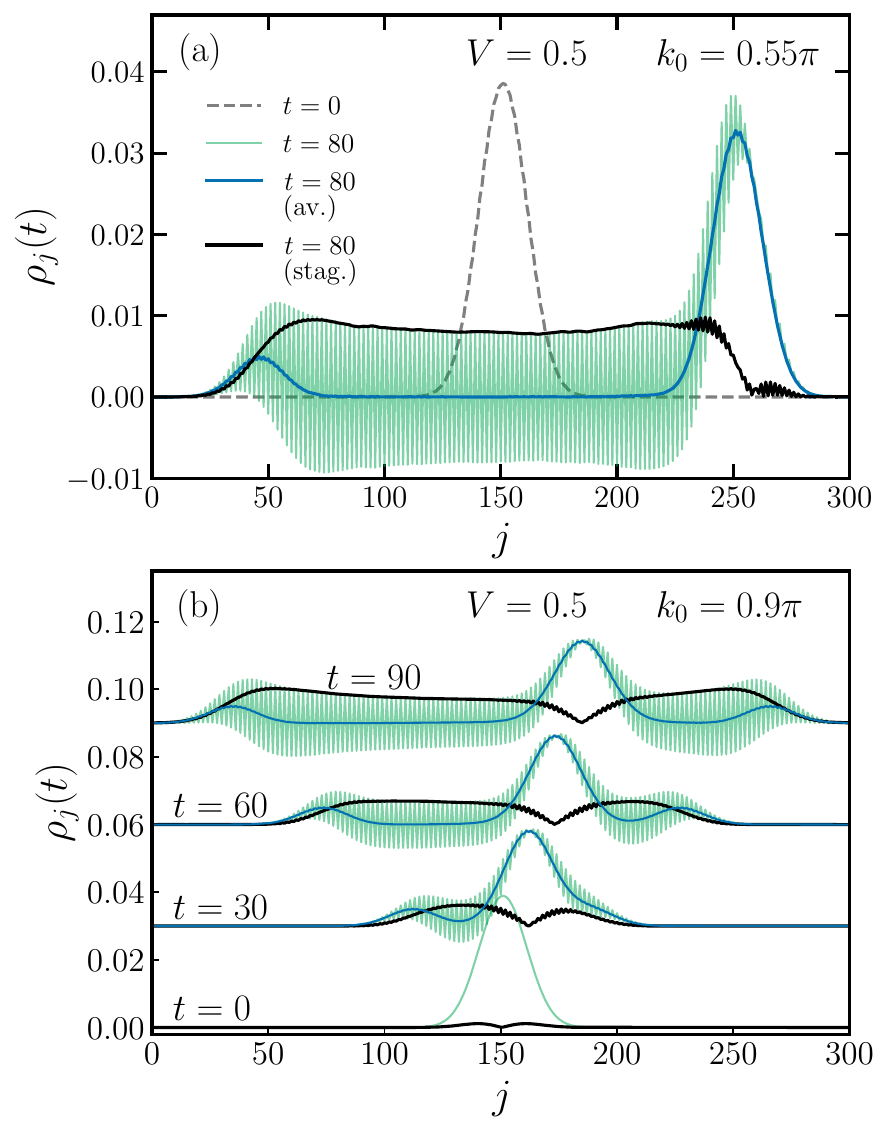}
		\par\end{centering}
	\caption{Snapshots of the density profile for (a) $k_{0}=0.55\pi$ and (b) $k_0=0.9\pi$, half filling and $V=0.5$. The green and blue lines represent the local charge excess before and after the two-site average, respectively. The black lines represent the two-site difference $\Delta \rho_j(t)$, see Eq. (\ref{diffrho}), which  selects the alternating part of the density profile.    \label{figsm2}}
\end{figure}

The numerical  results reveal  that in the interacting case the density profile  develops oscillations as soon as the counter-propagating humps start to split. This effect can be observed for the low- and high-energy regimes in Fig \ref{figsm2}(a) and (b), respectively.   To separate the  alternating  part from the smooth one, we evaluate the  absolute value of the difference between the local charge excess in neighboring sites, defining  \be
\Delta \rho_j(t)=\frac12|\rho_j(t) - \rho_{j+1}(t)|.\label{diffrho}
\ee
The result for $\Delta \rho_j(t)$ is represented by the black lines in Fig. \ref{figsm2}(b). The wave front of the alternating part moves along the smooth part of the fastest density humps, but the alternating part has a long tail that persists after the smooth part has propagated away. We stress that the alternating part does not contribute to the net charge being transported away from the region where the excitation was initially created. 

We can use the effective field theory to understand   how the oscillations arise only in the presence of  interactions.  The bosonized form of the density operator is   \cite{Giamarchi}\bea
n_j&\sim& \sqrt{\frac{K}{2\pi}}(\partial_x\varphi_L-\partial_x\varphi_R)\nonumber\\
&&+\frac1{2\pi\alpha}\cos\left[\sqrt{2\pi K}(\varphi_L-\varphi_R)-2k_Fx\right],
\eea
where $\alpha$ is a nonuniversal short-distance cutoff of the order of the lattice spacing. The second term is the staggered part that oscillates with momentum $2k_F$.

At low energies, we take $k_0\to k_F$ and  $c_j^\dagger \to e^{-ik_Fx}\psi^\dagger_R(x)+e^{ik_Fx}\psi^\dagger_L(x)$ in Eq. (\ref{exclat}). Setting $j_0=0$ in an infinite chain,  the expectation value of the staggered part in the time-evolved state becomes\bea
\rho^{\rm st}_j(t)&=&\frac{A^2e^{-i2k_Fx}}{4\pi \alpha} \int dx'dx''\, e^{-\frac{(x')^2}{2\sigma_0^2}}  e^{-\frac{(x^{\prime\prime})^2}{2\sigma_0^2}}e^{-i2k_Fx'} \nonumber\\
&&\times  \langle\psi^{\phantom\dagger}_L(x')e^{i\sqrt{2\pi K}[\varphi_L(x,t)-\varphi_R(x,t)]}\psi_R^{\dagger}(x'')\rangle+\text{c.c.}.\nonumber\\
&&
\eea
Bosonizing the fermion operators, we obtain
\bea
\rho^{\rm st}_j(t)&\sim & e^{-i2k_Fx} \int dx'dx''\, e^{-\frac{(x')^2}{2\sigma_0^2}}  e^{-\frac{(x^{\prime\prime})^2}{2\sigma_0^2}} e^{-i2k_Fx'}\nonumber\\
&&\times \langle \mc V_R(x';K-\nu) \mc V_R(x-vt;-K) \mc V_R(x'';\nu) \rangle \nonumber\\
&&\times \langle \mc V_L(x';-\nu) \mc V_L(x+vt;K) \mc V_L(x'';\nu-K)\rangle\nonumber\\
&&+\text{ c.c.}\label{integral}
\eea
where $\nu=\frac{1+K}{2}$. The three-point function for the chiral vertex operators has the form \bea
\langle \mc V_{R,L}(x_1;n_1) \mc V_{R,L}(x_2;n_2) \mc V_{R,L}(x_3;n_3) \rangle \nonumber\\
=\prod_{i<j} (x_i-x_j\pm i\alpha)^{n_in_j/K},
\eea
provided that the neutrality condition $\sum_{i}n_i=0$ is satisfied.   We then have \begin{widetext}
	\bea
	\rho^{\rm st}_j(t)&\sim & e^{-i2k_Fx} \int_{-\infty}^{\infty} dx''\, e^{-\frac{(x'')^2}{2\sigma_0^2}}  (x-vt-x''+i\alpha)^{-\frac{1+K}{2}} (x+vt-x''-i\alpha)^{\frac{1-K}2} \nonumber \\
	&&\times \int_{-\infty}^{\infty} dx'\, e^{-i2k_Fx'}e^{-\frac{(x')^2}{2\sigma_0^2}} (x'-x+vt+i\alpha)^{\frac{1-K}2}[(x'-x'')^2+\alpha^2]^{-\frac{1-K^2}{4K}}(x'-x-vt-i\alpha)^{-\frac{1+K}2}+\text{c.c.}.\label{toolong}
	\eea
\end{widetext}

From Eq. (\ref{toolong}) we can check that there are no oscillations in the noninteracting case.  Setting $K=1$ for $V=0$, the result simplifies to \bea
\rho^{\rm st}_j(t)&\sim & e^{-i2k_Fx} \int_{-\infty}^{\infty} dx''\, e^{-\frac{(x'')^2}{2\sigma_0^2}}  (x-vt-x''+i\alpha)^{-1}  \nonumber \\
&&\times \int_{-\infty}^{\infty} dx'\, e^{-i2k_Fx'}e^{-\frac{(x')^2}{2\sigma_0^2}} (x'-x-vt-i\alpha)^{-1}\nonumber\\
&&+\text{ c.c.}.
\eea
The integral over $x'$ vanishes because the integrand is an analytical function in the lower half plane. For $K\neq1$, the integrand in Eq. (\ref{toolong}) has branch cuts both above and below the real axis and the staggered part of the density operator acquires a nonzero expectation value. Physically, the branch points at $x'=x\pm vt$ are associated with excitations moving in opposite directions.

\subsection*{ Momentum independence at half filling} 

\begin{figure}
	\begin{centering}
		\includegraphics[width=7cm]{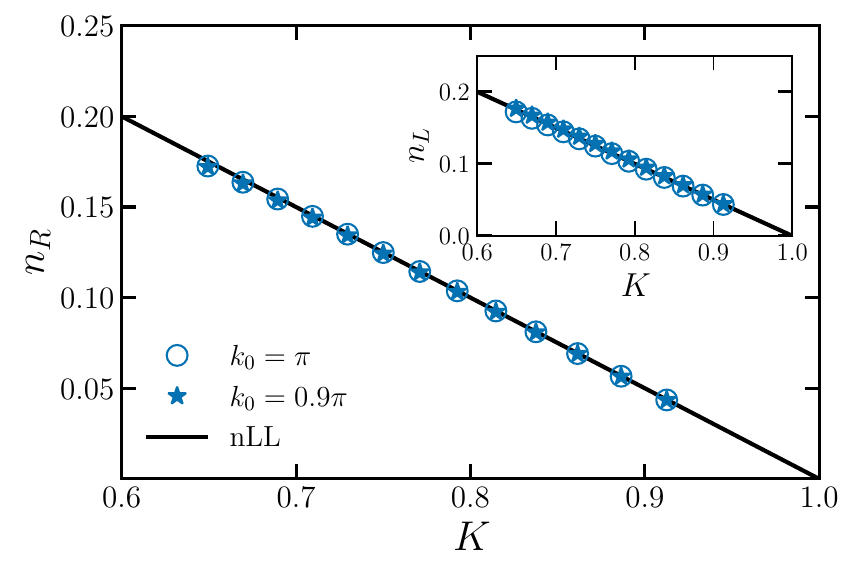}
		\par\end{centering}
	\caption{Charge carried by the right movers for $k_0=0.9 \pi$  and $\pi$. The charge of the left movers is shown in the inset. The solid lines  are the nLL predictions, $n_{R,L}=(1-K)/2$} \label{fig:chargehalf}
\end{figure}

To avoid the reflection of the wave packet at the boundaries,  the time reached in the tDMRG simulations is fixed by system size and the interaction. This limitation on the maximum time may prevent distinguishing the two right-moving humps. In Fig. \ref{fig:chargehalf}, we show the charges $n_{R,L}$ as a function of the LL parameter for two distinct values of $k_0$. In these cases, both right and left low-energy movers can be separately resolved and we observe an excellent agreement with the nLL prediction, $n_{R,L}=(1-K)/2$.

\begin{figure}
	\begin{centering}
		\includegraphics[width=7cm]{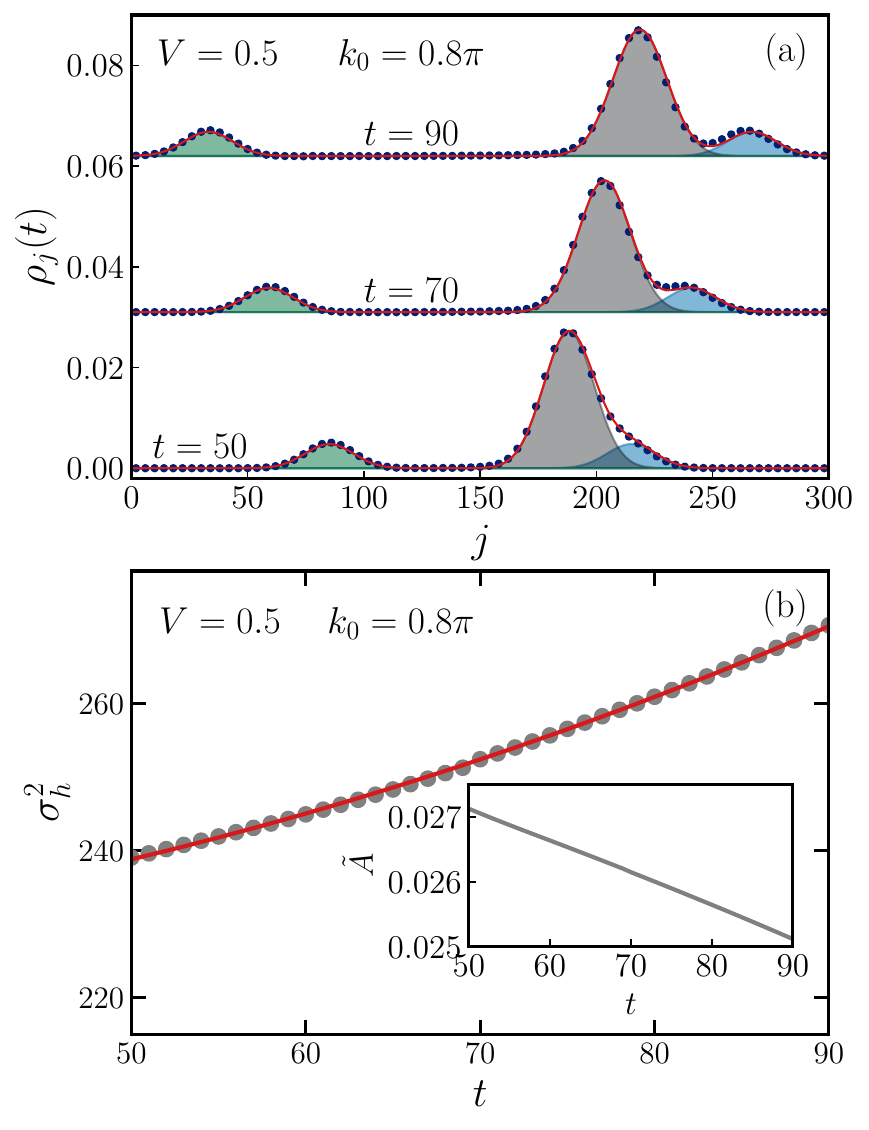}
	\end{centering}
	\caption{(a) Snapshots of the averaged density profile for $V=0.5$, $k_0=0.8\pi$ and $t=50,\,70$, and 90. The symbols are tDMRG results and the red lines fit our data considering Eq. (\ref{fitgauss}). The shaded regions represent the three Gaussian functions that are combined to produce the density profile. 
		(b) Time dependence of the fitting parameters $\sigma_l$ and $\tilde A$ (see the inset). The red line is the fit to our data using $\sigma_h^2(t)=a + bt^2$.
		\label{fig:combgauss}}
\end{figure}

Now, for the cases in which we cannot resolve the right-moving excitations, we analyze the dynamics by assuming that the density profile consists of two symmetric counter-propagating humps with velocity $v$ and a middle one with mean velocity $u=v\sin(k_0)$. We fit our results using the following equation

\begin{equation}
		\rho_j(t)=A \left[e^{-\frac{(j-j_0-vt)^2}{\sigma_l^2}}+e^{-\frac{(j-j_0+vt)^2}{\sigma_l^2}}\right]+\tilde A e^{-\frac{(j-j_0-ut)^2}{\sigma_h^2}},\label{fitgauss}
\end{equation}
where the amplitude $A$ and the variance $\sigma_l$ are obtained by fitting the left-moving wave packet. The amplitude $\tilde A$ and the variance of the high-energy wave packet are fitting parameters. In Fig. \ref{fig:combgauss}(a), we show combinations of three Gaussian functions that describe the time evolution of the charge wave packet. For $V=0.5$ and $k_0=0.8\pi$, the fixed parameters of the low-energy contributions are $\sigma_l=14.74$ and $A=0.0048$. The free fitting parameters $\tilde A$ and $\sigma_h$ depend on time and vary monotonically once the left movers can be discerned from the right-moving excitation; see Fig. \ref{fig:combgauss}(b). 

As aforementioned, in the free-fermion case, a Gaussian wave packet of a single particle that is strongly peaked around $k_0$ spreads in time as $\sigma^2(t)\approx \sigma_0^2(1+\frac{t^2}{m_0^2\sigma_0^4})$. In this context, by fitting our estimates of $\sigma_h^2(t)$ considering the function $\sigma_h^2(t)=a+bt^2$, we obtain $b=0.006$. Overall, we find that the combination of three Gaussian functions is consistent with our DMRG results.

\subsection*{Spectral decomposition of the initial excitation}

\begin{figure}
	\begin{centering}
		\includegraphics[width=7cm]{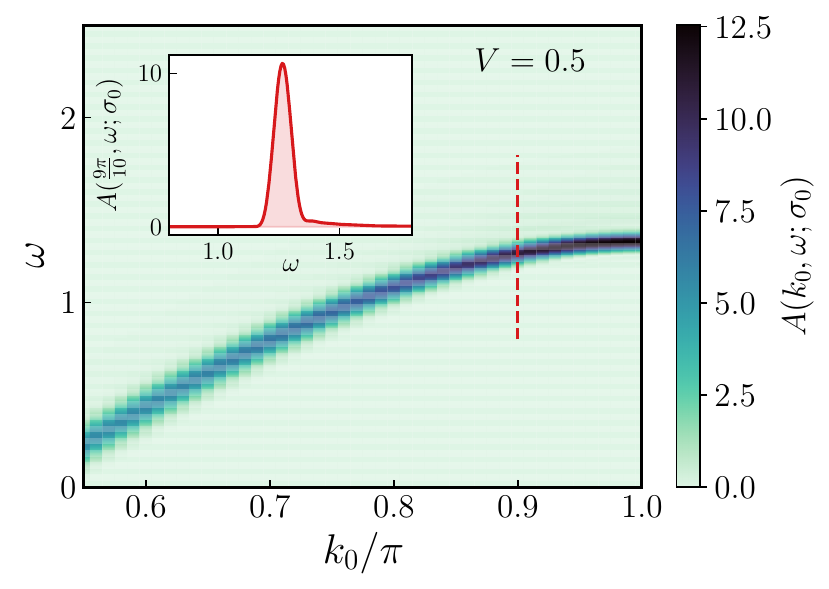}
		\par\end{centering}
	\caption{Function $A(k_0,\omega;\sigma)$ for $V=0.5$. The
		dominance of the high-energy particle excitations is unveiled by the single-peak pattern around $\varepsilon(k_{0})=-v\cos(k_{0})$. The inset shows a cut for $k_0=\frac{9\pi}{10}$. \label{fig:halffw}}
\end{figure}

In the main text, we discuss the spectral decomposition of the initial
excitation for quarter filling. In addition to the signature of one-particle
states, we observe the emergence of two-particle bound states, which
do not exist for the half-filling case \cite{Pereira2009}. Indeed,
the function $A(k_0,\omega;\sigma_0)$ is characterized by a single peak in the
vicinity of the high-energy particle excitations (see Fig. \ref{fig:halffw}).

\subsection*{Wave-packet dynamics at quarter filling}

\begin{figure}
	\begin{centering}
		\includegraphics[width=7cm]{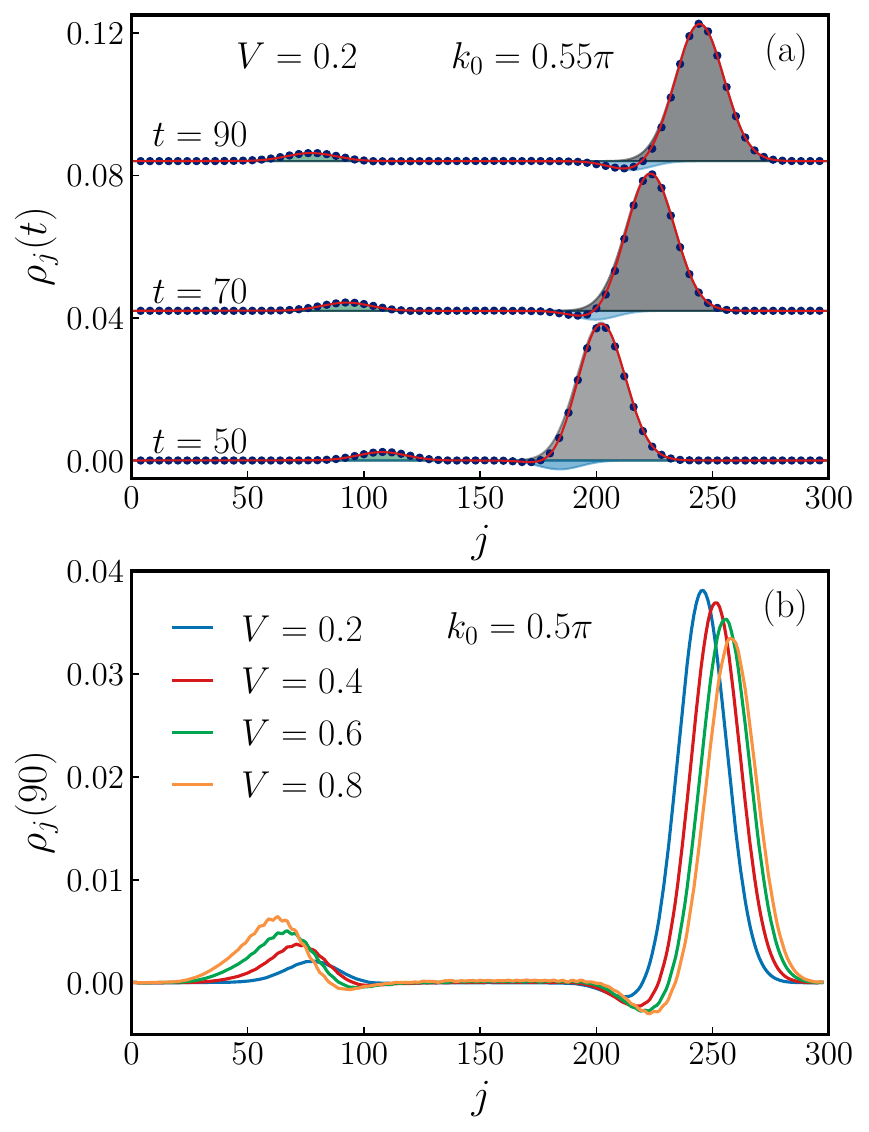}
		\par\end{centering}
	\caption{(a) Averaged density profile  for $k_{0}=\pi/2$, $t=90$ and distinct values of $V$. (b) Snapshots of the averaged density profile  for $V=0.2$ and $k_0=0.55\pi$. The time evolution of the wave packet can be described by the propagation of three Gaussian functions represented by the shaded regions on the plot, whose sum produces the red solid lines. The symbols are the DMRG results. \label{fig:lowdens}}
\end{figure}

In contrast to the half-filling case, where one-particle states dominate
the time evolution of the Gaussian excitation, additional high-energy
states play an important role at low densities. In this case, we
observe that the dynamics may harbor two-particle bound states and free
holes, whose fingerprint is observed in $A(k_0,\omega;\sigma_0)$. 
As discussed in the main text, the minimum value of momentum for which the aforementioned composite excitation exists is
\begin{equation}
	k\ge \pi + k_F - Q_{\text{bs}}.
\end{equation}
with $Q_{\text{bs}}=[\pi - 2 \arccos(V)]\left(1-\frac{2k_F}{\pi}\right)$. Thus, to observe the nLL prediction discussed in the main text, we must set appropriate  values of $k_0$ and $V$ such that \emph{only} single-particle excitations are created in the initial state. For instance, for $V=0.2$ and $k_0\lesssim 0.7\pi$, we observe that the late-time excitation comprises a three-hump structure, as predicted by nLL. In this regime, we can describe the evolution of the initial excitation using a combination of three propagating Gaussian wave packets. In Fig. \ref{fig:lowdens}(a), we show snapshots of the averaged density profile for $V=0.2$ and $k_0=0.55\pi$. The decomposition of the excitation into three Gaussian functions is represented by the shaded regions in the figure. The velocities $v=0.764$ and $u=1.06$ are the obtained via Bethe ansatz.  Note that the right-moving excitation is composed by one Gaussian function with a \emph{negative} amplitude that follows velocity $v$ and one hump with velocity of the high-energy particle with momentum $k_0$. It is worth pointing out that the negative charge carried by the low-energy right-moving hump is consistent with the nLL description, $n_R\propto1/(v-u)<0$.  In general, we found a good agreement between the combination of three Gaussian functions and the tDMRG results as long as we do not have composite excitations in the dynamics.

Finally, let us briefly discuss a case where we see signatures of the bound states, but we can still discern the left- and right-propagating excitations; see Fig. \ref{fig:lowdens}(b). Depending on $k_0$, the left-moving wave packet is not only composed by low-energy movers, but a combination of excitations with velocity close to $-v$. Indeed, the tail of the left wave packet in Fig. \ref{fig:lowdens}(b) suggests the existence of holes propagating with velocity close to the Fermi one.

\end{document}